\documentclass{ws-jai}
\usepackage[flushleft]{threeparttable}
\usepackage[utf8]{inputenc}

\usepackage{soul}
\usepackage[normalem]{ulem}
\usepackage{xcolor}

\newcommand{\sigTOT}{$\sigma_{TOT}$}

\begin{document}
\catchline{}{}{}{}{} 

\markboth{Belland et al.}{Focal Ratio Degradation}

\title{Focal ratio degradation for fiber positioner operation in astronomical spectrographs}

\author{Brent Belland$^{\text{1,5}}$, James Gunn$^2$, Dan Reiley$^1$, Judith Cohen$^1$, Evan Kirby$^1$, Antonio Cesar de Oliveira$^3$, Ligia Souza de Oliveira$^3$, Mitsuko Roberts$^1$, and Michael Seiffert$^{\text{1,4}}$}

\address{
$^1$Physics Department, Caltech, 1200 E. California Blvd., Pasadena, CA 91125, USA\\
$^2$Astronomy Department, Princeton University, 4 Ivy Lane, Princeton, NJ 08544, USA\\
$^3$Laboratório Nacional de Astrofisica, MCTI, Rua Estados Unidos, 154, Bairro das Nações, Itajubá, MG, Brazil\\
$^4$Jet Propulsion Laboratory, 4800 Oak Grove Drive, Pasadena, CA 91109, USA\\
$^5$bbelland@caltech.edu
}

\maketitle

\corres{$^5$Corresponding author.}

\begin{history}
\received{(to be inserted by publisher)};
\revised{(to be inserted by publisher)};
\accepted{(to be inserted by publisher)};
\end{history}

\begin{abstract}

Focal ratio degradation (FRD), the increase of light’s focal ratio between the input into an optical fiber and the output, is important to characterize for astronomical spectrographs due to its effects on throughput and the point spread function. However, while FRD is a function of many fiber properties such as stresses, microbending, and surface imperfections, angular misalignments between the incoming light and the face of the fiber also affect the light profile and complicate this measurement.
A compact experimental setup and a model separating FRD from angular misalignment was applied to a fiber subjected to varying stresses or angular misalignments to determine the magnitude of these effects. The FRD was then determined for a fiber in a fiber positioner that will be used in the Subaru Prime Focus Spectrograph (PFS). The analysis we carried out for the PFS positioner suggests that effects of angular misalignment dominate and no significant FRD increase due to stress should occur. 
\end{abstract}

\keywords{Focal ratio degradation, optical fiber, telescope, stress, misalignment.}

\section{Introduction} 

Optical fibers serve an important role in modern astronomical instrumentation. They are a key element in a number of current multi-object spectrographs (e.g., \citet{Smee2013,Lewis2002}), where they relay light from astronomical objects at the telescope focal surface to a remotely located spectrograph.  Optical fibers will find continued use in essentially all next-generation massively-multiplexed astronomical spectrographs. Projects under development 
include the Subaru Prime Focus Spectrograph (PFS) \cite{Sugai2015}, DESI \citep{Flaugher2014}, MOONS \citep{Cirasuolo2011}, 4MOST \citep{deJong2012}, and WEAVE \citep{Dalton2012}, among others. 
This paper is motivated by focal ratio degradation (FRD) testing for the Subaru PFS, utilizing the same fiber and fiber positioning system that will be used in the PFS.

\begin{figure}[!htb]
\center{\includegraphics[width=12cm]
{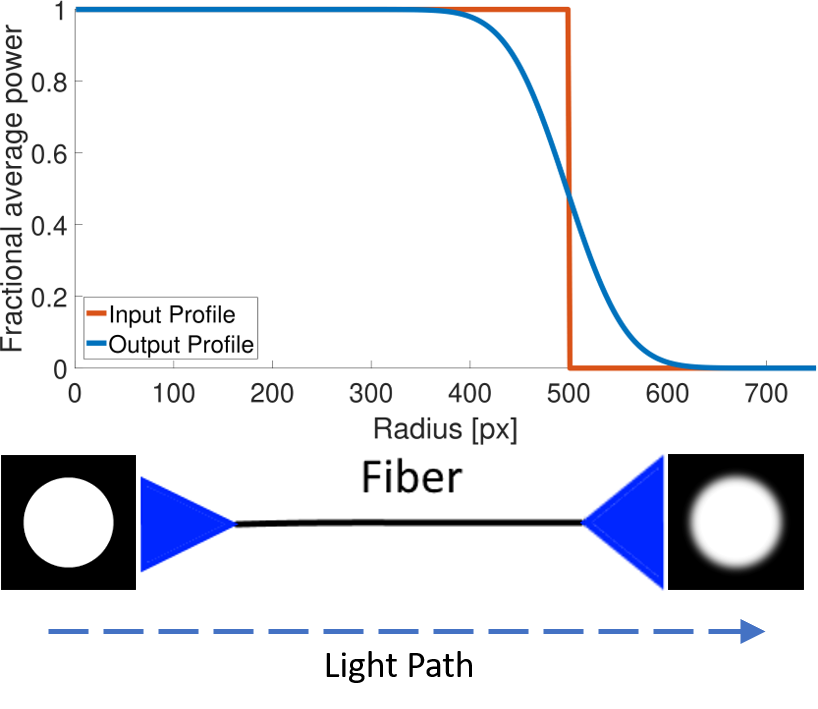}}
\caption{\label{fig:FRDschematic} Simulated effect of FRD on a uniform tophat profile. Above, the radial profile of the image when input and output from the fiber is shown. Below is a two dimensional image that demonstrates the effect FRD has on the point spread function has on FRD. Notably, an instrument that could only accept up to the initial tophat would lose throughput due to the FRD-induced increased profile size. Even if the instrument can accept faster inputs, increasing FRD causes increasing illumination of the outer parts of the 
pupil, which for the fast spectrograph cameras in all the projects
discussed in the introduction brings increasing aberrations and degradation of the
point spread function.}
\end{figure}

FRD is the decrease of the focal ratio of a beam traversing a fiber \cite{Ramsey1988}. The light power distribution thus varies due to FRD after transmission through a fiber, causing issues for subtracting night sky lines in fiber-fed spectrographs in astronomical implementations \cite{Clayton1989,Bolton2010}. These issues must be addressed in order to limit the sky subtraction residuals to less than 0.5\% of the sky continuum, as is the goal for the PFS \citep{Tamura2016}. The FRD of the fiber is a function of many components, potentially varying as the fiber experiences different stresses during positioning and telescope operation. FRD in the PFS has a dynamically changing effect on the point spread function, as will be discussed below.
Furthermore, FRD can cause a decrease in throughput if it causes some light to scatter to a focal ratio that is smaller than the acceptance focal ratio of the spectrograph. An example of how FRD can affect a beam is demonstrated schematically in Figure~\ref{fig:FRDschematic}.

The causes of FRD include any effect that can scatter the light, including polishing irregularities at the fiber surface, microbending introducing irregularities to the core/clad interface, diffraction, bending, twisting, and squeezing among other causes \cite{Clayton1989,Oliveira2005}. It should therefore be minimized to diminish the throughput loss and quantified to understand its effect on the light's point spread function.

However, the key to precise atmospheric background subtraction for PFS is the accurate modeling of the point spread function, not just knowledge of the FRD. This in turn requires an accurate model of the intensity as a function of angle when light exits the fiber and enters the spectrograph. In particular, any change with the instrument configuration, such as the position in the focal plane or the position of individual fibers within their patrol region that affects the PSF will need to be determined.  With this in mind, we conducted a series of experiments to evaluate changes in FRD in a fiber due specifically to its fiber positioning hardware which we call a Cobra. For these purposes, we wish to know not only the net effect, but also the components due to applied fiber stress (that is, the bending and twisting inside the Cobra due to its motion), angular misalignments (both static and dynamic). The repeatability and predictability of the FRD is also of foremost importance.

There are two primary methods to measure FRD: a cone test,also known as a solid angle test, and a ring test, also known as a collimated beam test \cite{Haynes2011,Yan2018}. The cone test involves a uniform beam of known solid angle that fills a fiber. The output profile is measured and typically the enclosed energy at varying f-ratios is found \cite{Oliveira2005,dosSantos2014}; relative enclosed energies are compared to determine FRD. The cone test is able to probe the whole input cone of the fiber, but it is sensitive to misalignments. In contrast, the ring test involves a beam input at a known angle into a fiber, quantifying FRD by the FWHM of the resulting ring output \cite{AllingtonSmith2013,Finstad2016}. While ring FRD can be quantified by its angular size, it is restricted to a small range about the input f number and characterize the light power distribution over the entire fiber outside of this range. Comparison of results between these two tests is complicated by the fact that the measurement outputs are different.

In this paper, a method of determining absolute FRD from the cone test is proposed. Such a system would be able to probe the whole input range of a measured fiber, being sensitive to input profile variations in a single test, while still quantifying the FRD in a way comparable to the ring test. In particular, the misalignment component of the cone profile can be extracted, yielding both fiber FRD and angular misalignment at the same time.

This paper is structured to discuss the FRD extraction model considered here in Section~\ref{Model} and then hardware setup in Section~\ref{hardware}. Afterward, various forms of stress (bending, squeezing, and twisting), and thus potentially of FRD, are measured in a Polymicro fiber in Section~\ref{stresses}, and angular misalignment in a fiber is interpreted using the cone test in Section~\ref{AngMis} to compare to the angular component of the model. Finally, each of these components is considered in the fiber positioning system, the Cobra, which will be used in the Subaru PFS, to determine the relative magnitude and contributions of stress vs. angle in Section~\ref{Cobra}.

\section{Model}\label{Model}

FRD is the effective smearing of light due to imperfections of the fiber core/clad interface from fabrication and from stresses such as from bending, as well as from fiber end face preparation. Assuming these defects are scatterers of a random angle and over the length of the fiber many such defects are encountered, by the Central Limit Theorem, the angle scattered should roughly be describable by a Gaussian distribution with standard deviation $\sigma$. In practice, \citet{Haynes2011} found using the ring test that modal diffusion due to effects such as microbending behaved in a Gaussian way, aperture diffraction could be approximated as a Gaussian, and end face surface roughness could be modeled by a Gaussian or Lorentzian profile with Lorentzian component decreasing with smoother fiber surface. In this paper, the Gaussian effects are assumed to dominate FRD.

To first order we wish to characterize FRD by the $\sigma$ of a Gaussian that, when convolved with a tophat input light distribution, yields the observed output profile. Without any other outside influences, this $\sigma$ would characterize FRD, but angular misalignments can confound this analysis. Let the $\sigma$ value due to FRD be $\sigma_{FRD}$ whereas $\sigma_{TOT}$ will be used to characterize the $\sigma$ that would create an equivalent width of the entire dropoff of power over angle. We can find $\sigma_{TOT}$ directly from analyzing a fiber's output profile, but $\sigma_{FRD}$ can also be extracted due to the different effects FRD and angular misalignment have on the profile. In the case of no angular misalignment, $\sigma_{TOT} = \sigma_{FRD}$.

To verify the initial premise of this model, namely that scattering from FRD can be quantified with a value for $\sigma$, previous work must be considered. FRD due to scattering of light causing light coupling to higher modes within the fiber has been analyzed by \citet{Gloge1972}, who developed a model to describe the power distribution $P$ of light from a fiber of length $L$ due to microbending, as a function of angle from normal incidence $\theta$:

\begin{equation}
\frac{\partial P}{\partial L} = -A\theta^2 P +\frac{D}{\theta}\frac{\partial}{\partial\theta}\left(\theta\frac{\partial P}{\partial\theta} \right)\label{eq:GlogePower}
\end{equation} with $A$ and $D$ corresponding to absorption and coupling coefficients respectively. 

This power distribution PDE was solved by \citet{Gambling1975} for a plane wave at an angle of incidence $\theta_i$:

\begin{equation}
P(\theta)\simeq \frac{1}{bL} \exp\left({-\frac{\theta^2+\theta_i^2}{4DL}}\right)I_0\left(\frac{\theta\theta_i}{2DL}\right)\label{eq:GamblingPlane}
\end{equation} for $b = 4\sqrt{AD}$ and $I_0$ representing a Bessel function of zeroth order, for $bL \ll 1$. Effectively, this predicts the ring test output. Notably, this expression asymptotically approaches Gaussian behavior in the limits $(\frac{\theta\theta_i}{2DL}) \ll 1$ and $(\frac{\theta\theta_i}{2DL}) \gg 1$, both with standard deviation 

\begin{equation}
\sigma = \sqrt{2DL}.\label{eq:GlogeSigma}
\end{equation}

In order to relate the results of Gambling et al. to the cone test, due to the simple nature of the plane wave, it is possible to integrate to create an input profile for a cone of light; integrating over an input profile $G(\theta,\theta_i)$ yields an output power $F$ of \begin{equation}
F(\theta) = \int_0^{2\pi}\int_0^{\pi}G(\theta',\phi')P(\theta,\theta')\sin(\theta')d\theta'd\phi'.\label{eq:FancyInt}
\end{equation}Assuming a uniform cone-like beam with rotational symmetry about the fiber axis, the overall output power $F$ from the fiber then becomes 
\begin{equation}
F(\theta) \propto 2\pi\int_0^{\theta_i}P(\theta,\theta')\sin(\theta')d\theta'.\label{eq:SimpleInt}
\end{equation}This integral approximates a normal error function with the same $\sigma$ as the one corresponding to that of the ring test. 

Thus, the FRD-only model aligns well with a $\sigma_{FRD}$ characterizing its output, just as utilized in this model. 

However, in general, there is more than FRD affecting the profile; $\sigma_{TOT} > \sigma_{FRD}$ due to angular misalignments. If the chief ray is not normal to the fiber face, a tophat input does not appear to be a tophat at the fiber. Instead, the optical system is at a nonzero angle relative to the fiber axis. 

This has the effect of convolving each point in the tophat with a ring of size corresponding to its angular misalignment. That is, the effect of angular misalignment is not to convolve the profile with a Gaussian as microbends do; by analyzing the shape of a profile dropoff, the effect of angular misalignment can be separated from that of microbend FRD. In 2D, such an angularly misaligned profile from rings is difficult to calculate analytically; however, in 1D, the convolution kernel of a profile with an angular misalignment $a$ would behave as 

\begin{equation}
k(\theta) \propto \frac{1}{\sqrt{1-(\frac{\theta}{a})^2}}\label{eq:Convkernel}
\end{equation} Note how this convolution kernel is bimodal at $\theta = \pm a$, yielding a dropoff distinct from a gaussian convolution. Such behavior has a distinct effect on the point spread function compared to FRD and must be considered independently.

To first order, the net width of the dropoff, \sigTOT, is then the sum of the rms angular misalignment kernel width (of $\frac{a}{\sqrt{2}}$) and the Gaussian kernel width (independent of misalignment, called $\sigma_{FRD}$). That is, $\sigma_{TOT}^2 \approx \sigma_{FRD}^2+\frac{a^2}{2}$. A more thorough analysis of the convolved functions approximates an analytic FRD $\sigma_{TOT}$ of

\begin{equation}
\sigma_{\text{TOT}}^2 \approx \sigma_{\text{FRD}}^2\left(1 + \frac{1+A^2}{2+1.5A^2}A^2\right)\label{eq:JimAngle}
\end{equation} where $A=\frac{a}{\sigma_{\text{FRD}}}$. This estimate of $\sigma_{TOT}$ varies by $<1\%$ from the numerical integration result for angular misalignments $A\le 6$. 

\begin{figure}[!htb]
\center{\includegraphics[width=16cm]
{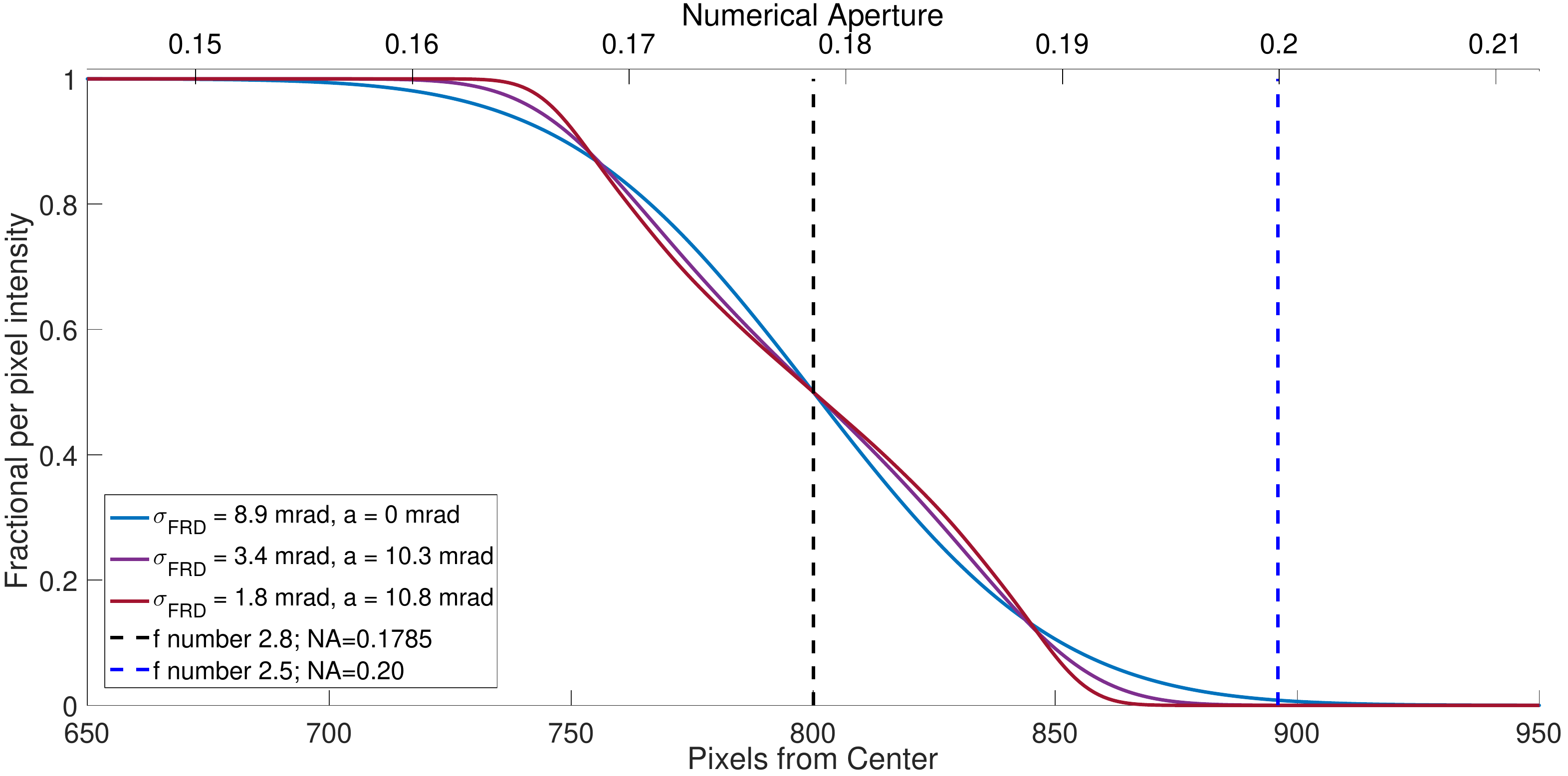}}
\caption{\label{fig:ExpAngMisExplain} Dropoff in pixel intensity for the angular misalignment model with uniform input beam of f number 2.8 (numerical aperture of 0.1785). The effect on the profile for the same net width $\sigma_{TOT}=8.9$ mrad are shown for varying angular misalignments, which corresponds to a different gaussian contribution from FRD (large angles are shown to emphasize the effect of angular misalignment). Also shown is the f number of 2.5 corresponding to the spectroscope cutoff. The light lost at the cutoff varies at a given $\sigma_{TOT}$ for different angular misalignments.} 
\end{figure}

Figure~\ref{fig:ExpAngMisExplain} visualizes how angular misalignment may affect the profile of the dropoff using a model of a uniform input profile, for a given $\sigma_{TOT}$. Notably, the error function dropoff from a purely Gaussian convolution becomes bimodal with large angular misalignments, and the effective width of the profile for a given FRD increases with angular misalignment. Note that holding $\sigma_{TOT}$ constant for a range of angular misalignments allows for a comparison of angular misalignments, but for a fixed $\sigma_{FRD}$ angular misalignments increase the width of the dropoff.

Also from Figure~\ref{fig:ExpAngMisExplain}, note that from the model of a uniform beam with f number 2.8 (that is, numerical aperture, or NA, of 0.1785), the amount of light cutoff at f number 2.5 (NA of 0.2) can also be seen and the light cutoff can be determined. At 8.9 mrad of $\sigma_{TOT}$ and no angular misalignment, only 0.03\% of the light is cut off, whereas for comparison, at 20 mrad of $\sigma_{TOT}$ and no angular misalignment, about 1.9\% of the light is blocked.

\begin{figure}[!htb]
\center{\includegraphics[width=16cm]
{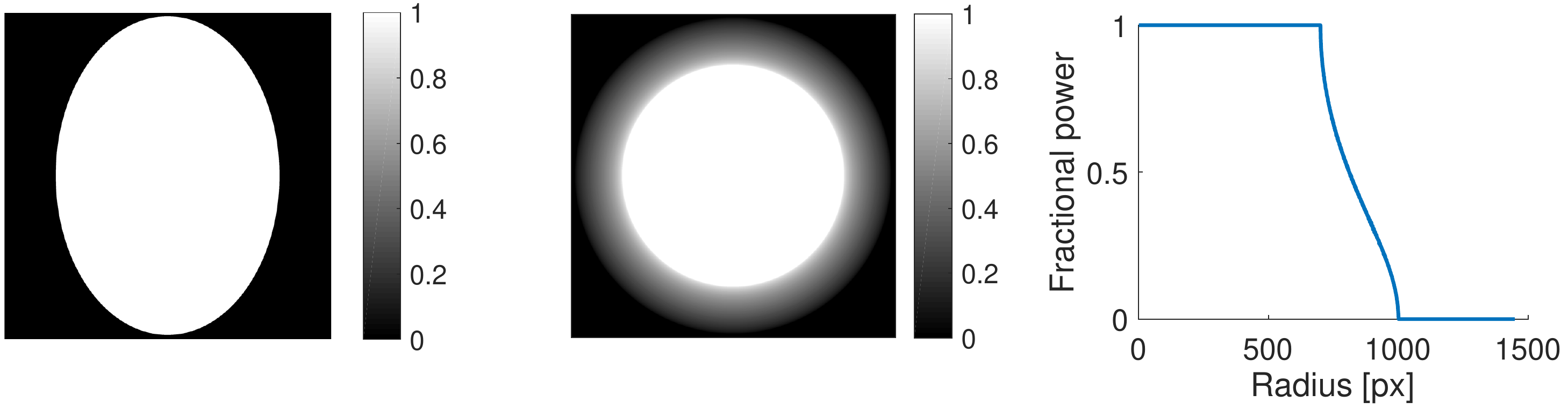}}
\caption{\label{fig:Oval} Simple example of an ellipse profile (left) that to first order approximates a tilted circle. A figure demonstrating what a 2D profile with the rotationally averaged profile is shown in the middle panel, with its 1D profile from the center shown on the right. Notably, the dropoff in the 1D profile does not appear to behave like an error function; instead, it is similar to an image convolved with a ring. This suggests how our model could be related to Gloge's power distribution model.}
\end{figure}

Again, this model should be related back to Gloge's modal diffusion theory; angular misalignments can be factored in with an appropriate choice of $G(\theta',\phi')$ using equation~\eqref{eq:FancyInt}. Detailed analysis is left for future work; however, the feasibility of this paper's model can be tested using a simple $G(\theta',\phi')$. A circle in a plane tilted at an angle with respect to a viewer appears to be an ellipse. Thus, consider the profile of an ellipse for $G(\theta',\phi')$. Because only $G$ depends on $\phi'$ in Equation~\eqref{eq:FancyInt}, we can substitute it with a rotationally averaged over $\phi'$ function of $\bar{G}(\theta')$; the 1D profile of a rotationally averaged ellipse has an approximately bimodal distribution similar to a step function convolved with the kernel in equation~\eqref{eq:Convkernel}. An example of such a convolution is shown in figure~\ref{fig:Oval}. Overall, the simplified model is consistent with FRD results in the literature and allows for a simple quantification of angular misalignment that is not unreasonable.

\section{Hardware}\label{hardware}

    In order to measure FRD well, the input illumination into the fiber must be well-characterized, and the camera must have enough precision to determine small variations in the output profile. This experiment is designed to compactly measure FRD variations in a fiber illuminated with uniform intensity up to a given numerical aperture. Furthermore, the experiment is designed so that the illumination is uniform over a spot of a $\gtrsim 9.5$mm size, putting a less strict requirement on fiber and illuminator alignment. But the primary purpose of this design is to enables FRD measurement of a fiber throughout its range of motion in a Subaru fiber positioner without constant realignment; as stated in Section~\ref{Cobra}, the fiber positioner patrol region size is 9.5mm and fits within the illumination spot size.

\begin{figure}[!htb]
\center{\includegraphics[width=16cm]
{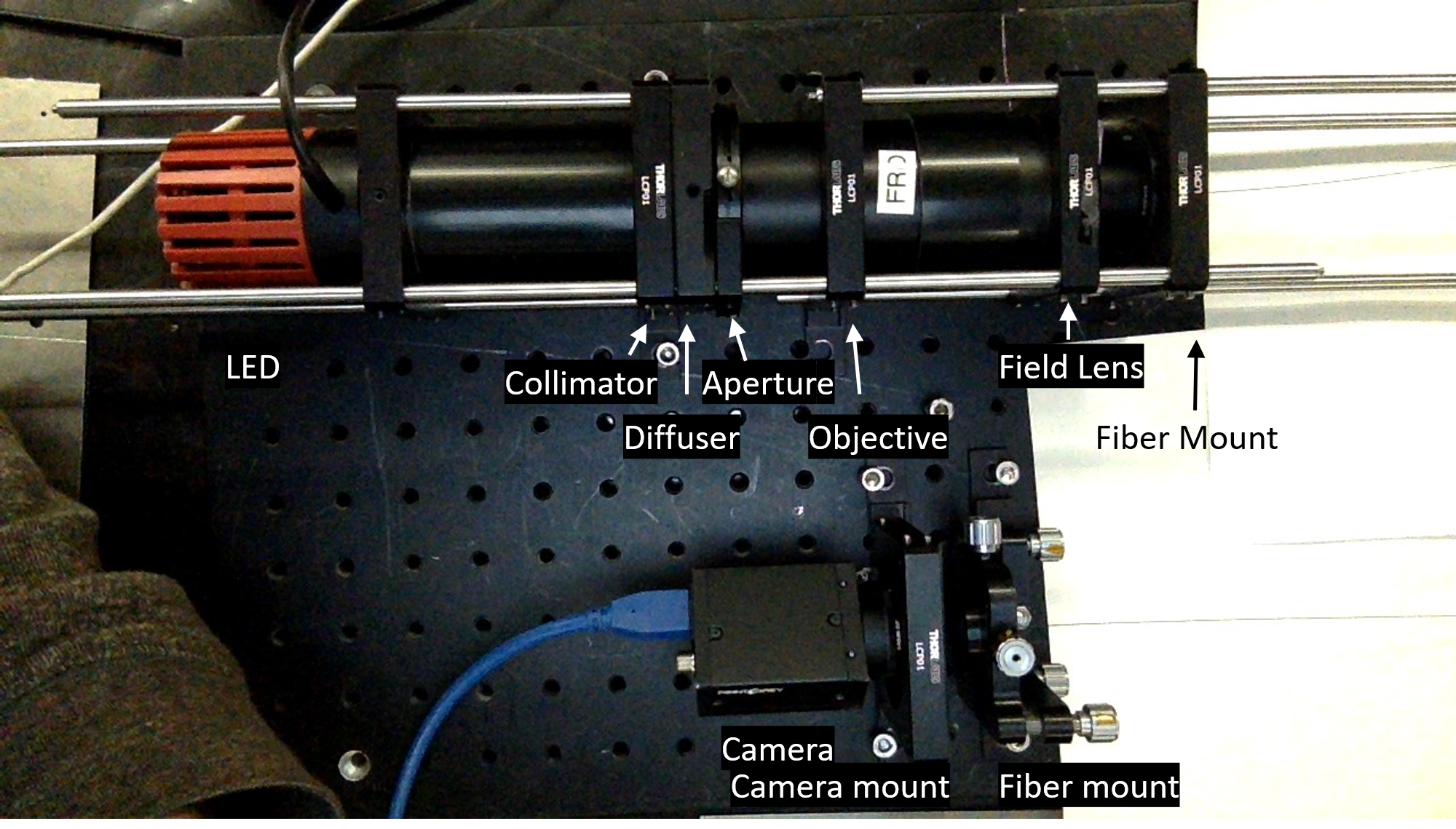}}
\caption{\label{fig:Apparatus} Picture of the main camera and illuminator setup of the experiment, with fiber extending off the right side of the image. Components of the illuminator and camera setup are labeled.}
\end{figure}

\begin{figure}[!htb]
\center{\includegraphics[width=16cm]
{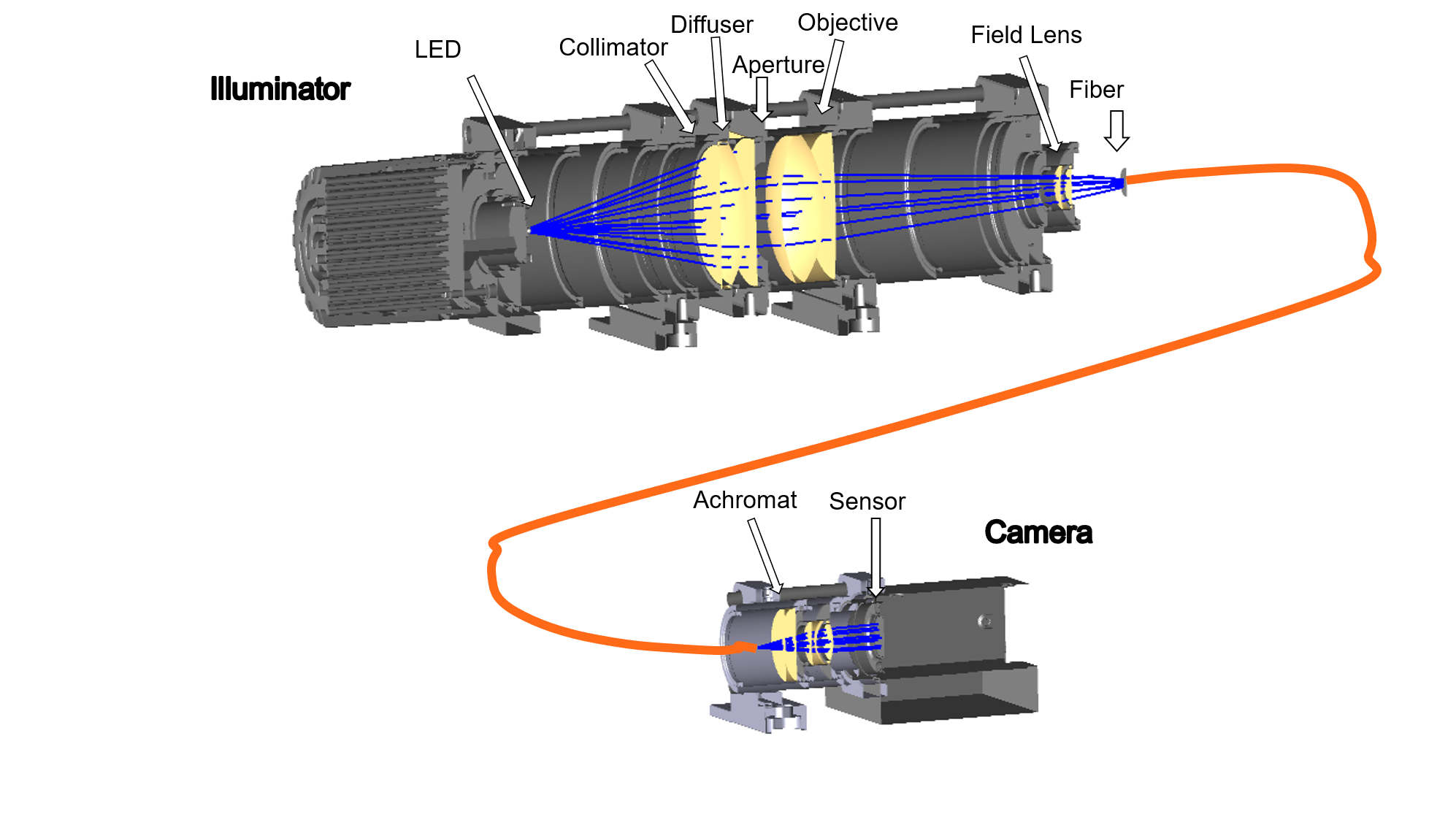}}
\caption{\label{fig:ApparatusZ} Cutout schematic view from above of the apparatus used in the experiment, with blue rays outlining the optical path through the system. The illuminator is at the top of the figure consisting of the LED through the field lens, while the camera is at the bottom with the achromat and sensor. The fiber in a Cobra fiber positioner was also fitted with a microlens to map a f/2.2 beam to an f/2.8 beam. The illuminator and camera were mounted securely to an optical breadboard and the camera was covered with a shroud during measurement.}
\end{figure}

The illuminator system generates a 10mm diameter telecentric image with uniform pupil illumination.  Light from a 450nm LED source is collimated by a 100mm focal length, 50mm diameter planoconvex lens.  An aperture stop is formed by an EDC-15-08160-A engineered diffuser from RPC Photonics, which uniformly distributes the collimated beam into a $15^\circ$ cone.  At the aperture stop, an 18-vane adjustable iris provides a round aperture of adjustable diameter.  Light is focused to the image plane with a 100mm focal length, 50mm diameter doublet, located near the aperture stop. A 100mm focal length, 15mm diameter plano-convex singlet is used as a field lens, ensuring the illumination is  across the image plane. All optics were mounted into an off-the-shelf cage system, and baffling was used to shield the illuminator from external light. The experimental illuminator and camera are shown in Figure~\ref{fig:Apparatus} and in a Zemax schematic in Figure~\ref{fig:ApparatusZ}.

The illumination color was chosen to be approximately monochromatic to avoid concerns about FRD variations with wavelength, because there is literature indicating that FRD effects increase with increasing wavelength. In particular, Gloge's coupling coefficient $D$ (see Section~\ref{Model}) is proportional to $\lambda^2$, causing the predicted FRD of his model (equation~\eqref{eq:GlogeSigma}) to be proportional to wavelength, and there are groups that have experimentally found such a dependence (e.g., \citet{Poppett2007}). As a result, FRD in the red to near infrared bands measured in the PFS may be larger than the values calculated here. Wavelength dependence with FRD is not fully established, however; for instance, see also \citet{Murphy2008}. 

The light from the illuminator is incident upon a Polymicro fiber with model number FBP127165190 (part number 1068020148) indicating a FBP fiber with a 127 $\mu$m core diameter, 165 $\mu$m clad diameter, and 190 $\mu$m buffer diameter; the same type of fiber as will be used in the PFS. FBP fiber is a low hydroxyl (-OH) fiber well-suited for astronomy, designed to accept broadband illumination from near infrared to ultraviolet; additional measurements on wavelength-dependent throughput and FRD of this fiber are in \citet{dosSantos2014}. The fiber end is terminated and either fitted with a zirconium ferrule or, when in a Cobra fiber positioner, fitted with a microlens in the Cobra fiber arm (See section \ref{Cobra}). The microlens takes an f/2.2 beam at the edge of the acceptance range in the fiber and increases its f number to f/2.8. The f/2.8 beam is slow enough to comfortably be accepted by the f/2.5 acceptance cone of the spectrograph in the final system, minimizing throughput losses. While the microlens may have its own effect on FRD affecting the final image, this was not observed in our tests.

The image from the fiber is projected onto a CMOS camera system, with a 15mm diameter, 25mm focal length achromatic lens to focus the fiber light onto the camera sensor. Although the camera was mounted onto an optical breadboard, is was not aligned with the fiber end by cage system. Alignment was completed visually with a Newport LPV-1 5-axis mount that held the fiber end (terminated with a zirconium ferrule) based on the output image from the camera. During operation the camera mount was covered with a black shroud to prevent stray light from entering the camera.

Image data was analyzed via a Matlab program interfacing with the camera. The center of the profile was determined after fitting for its center after subtracting out a background to remove scattered light and detector offset. Additionally, flats were taken for the camera pointed at a white background. While analysis was done without the flat correction initially, a later flat was used to determine the camera response and vignetting. This correction had a minimal effect (on the order of 0.1 mrad) on the calculation of $\sigma_{TOT}$. The center of the circle was found by calculating ``white'' and ``dark'' values by fitting Gaussians to histograms of number of instances for each brightness occurrence, then finding all points with 30\%, 50\% and 70\% of the ``white'' minus ``dark'' values, fitting each contour to the equation for a circle. Afterward, the azimuthal average of points at the same distance rounded to the nearest pixel from the center was taken and combined into a 1D profile, from which the FRD was analyzed. Small shifts on the order of a pixel could affect FRD measurements by 0.1 mrad, a nontrivial variation, so center finding was analyzed carefully. To verify that the center was indeed found, in initial images the center was tested by shifting the center within a 5px x 5px square, with the effects on the 1D profile (and in particular, \sigTOT) analyzed.

\begin{figure}[!htb]
\center{\includegraphics[clip,width=16cm]
{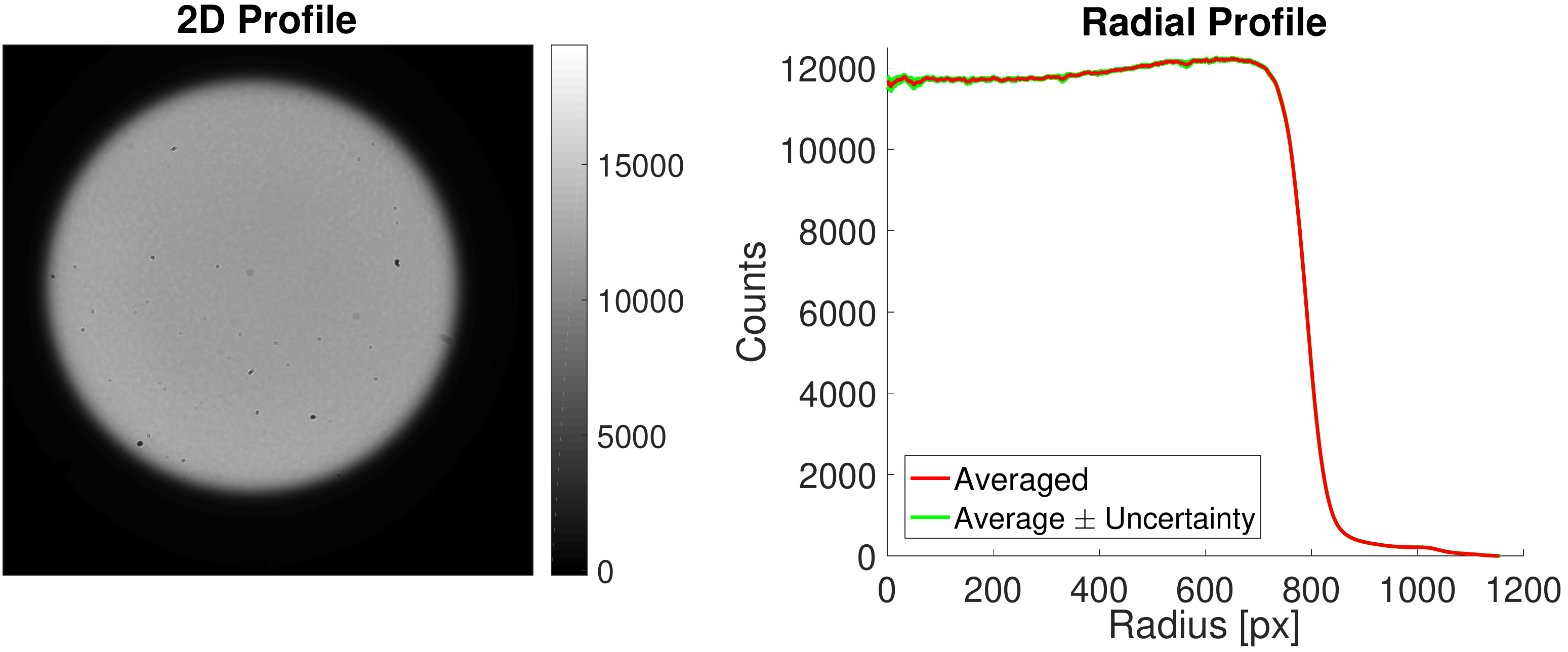}}
\caption{\label{fig:pipeline} Example data analysis image output from 2D image (left) scaled according to counts at the camera to radial profile extracted from the fitted center. The darker spots visible in the image were dust particles and are not intrinsic to the profile, though radially averaging minimizes the effect this has on the data.
}
\end{figure}

\begin{figure}[!htb]
\center{\includegraphics[clip,width=16cm]
{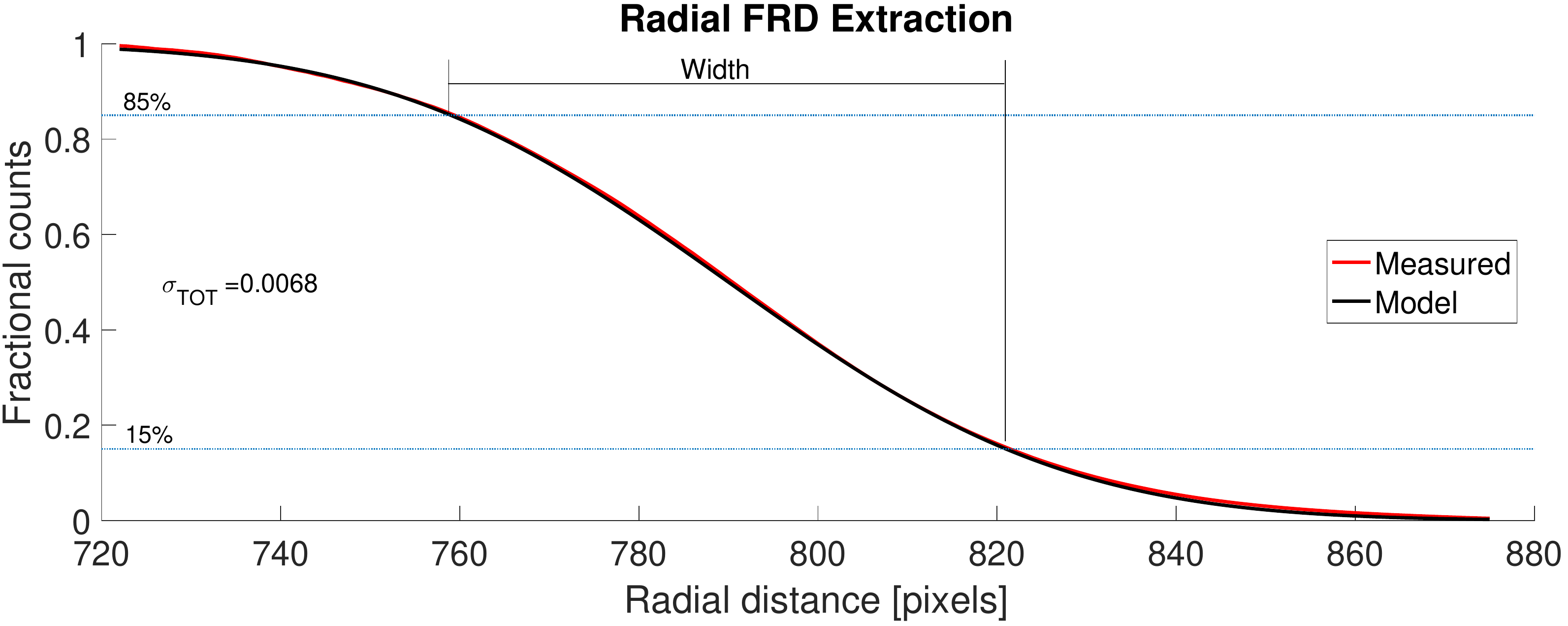}}
\caption{\label{fig:pipelinetwo} A zoomed-in image of the radial profile around the region of interest, the dropoff. A model normal error function calculated from the difference between the 85\% and 15\% brightness levels (dashed lines) is in good agreement with the observed profile. The width, as indicated in the figure, can be extracted and is proportional to $\sigma_{TOT}$.}
\end{figure}

To characterize the dropoff of the profile due to FRD, we found the width ($\sigma_{TOT}$) of a 2D Gaussian that, when convolved with a uniform tophat profile, was the same width as the observed dropoff (section~\ref{Model}). An example of this procedure is demonstrated in Figure~\ref{fig:pipeline}, which shows the 2D profile and radial profile, and fit for $\sigma_{TOT}$ depicted in Figure~\ref{fig:pipelinetwo}. In practice, the width of the dropoff (from 85\% of max brightness to 15\% of max brightness) was calculated and scaled to $\sigma_{TOT}$. Fitting for the 85\% and 15\% allows for a fit that is less biased from small deviations from a normal at large angles due to scattered light, such as due to the small bump at large radius visible in Figure~\ref{fig:pipeline}. Furthermore, this fitting method is more meaningful when angular misalignment is more prominent; each profile for a variety of angular misalignments is normalized to the width of the dropoff between the 85\% and 15\% brightness levels, as described in more detail in section~\ref{AngMis}.

Measurement of the widths of the profile dropoffs are made in pixels and then converted into equivalent angular size. However, this means that measurement precision is dictated by pixel size; in practice a pixel corresponded to approximately 0.1 mrad. Multiple images from the same experimental state had dropoff width varying with $\pm 1$ pixel, corresponding to a $\pm 0.1$ mrad uncertainty from width calculation.

An uncertainty in the calculation in this method was caused by the calculation of the 85\% and 15\% brightness radial locations. In particular, the uncertainty for these points was taken to be half the range of radii with the 85\% and 15\% values within one standard deviation of the average value (scaled for counting statistics with number of uncorrelated data points, respectively; see section 1.2.1).

\begin{figure}[!htb]
\center{\includegraphics[clip,width=16cm]
{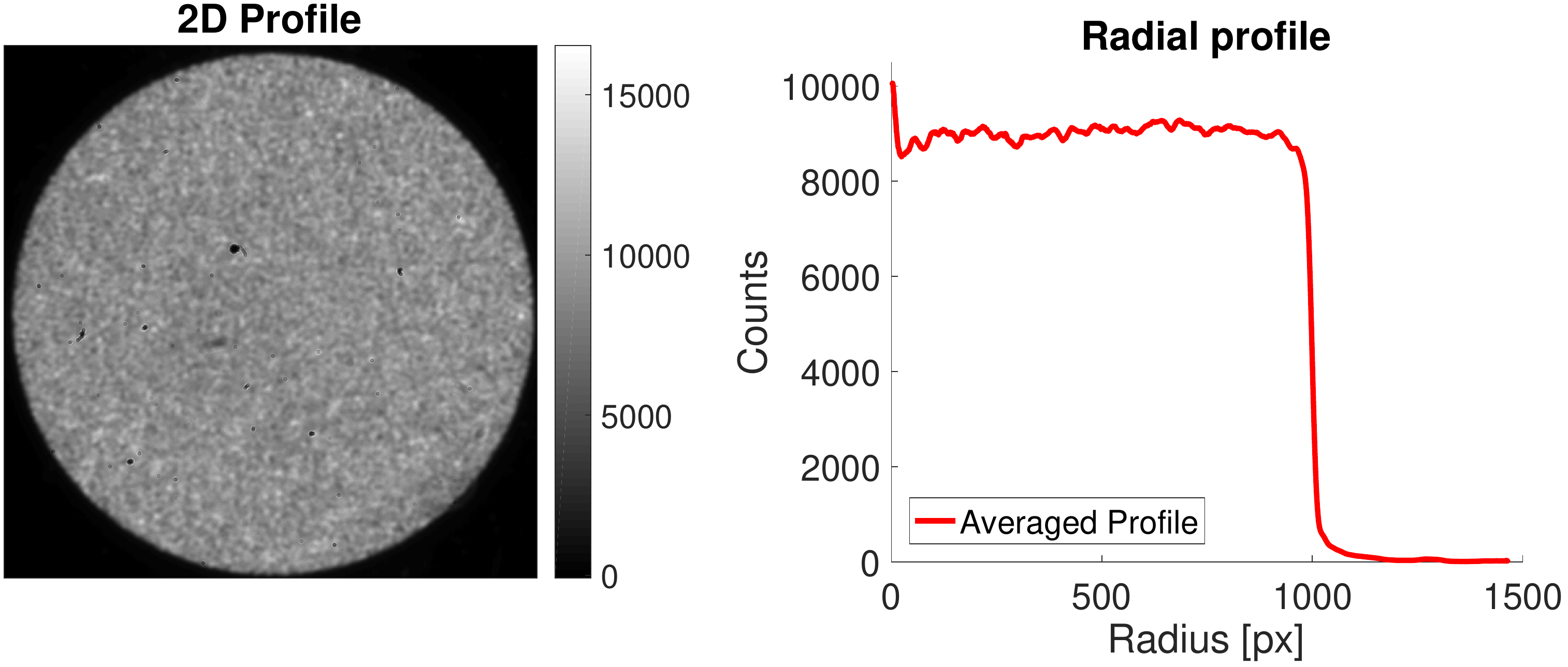}}
\caption{\label{fig:pinhole} 
Measurement with the fiber replaced by a pinhole. This measurement represents the instrumental limit for the low FRD condition. The correlated speckles in the 2D profile are likely due to fabrication variations across the surface of the diffuser. The width of the dropoff in the radial profile is much lower than the dropoff measured from fibers; the nonzero width may indicate a systematic - though small - overestimation of FRD.
}
\end{figure}

The basic data analysis was verified with a 100 $\mu$m pinhole replacing the fiber to visualize the input to the fiber. The resulting image is shown in Figure~\ref{fig:pinhole}. Notably, the pinhole behaves like a zero length fiber, so it should have a sharp dropoff, whereas there is a nonzero width dropoff actually observed. 
This effect is expected to add in quadrature with the FRD to give a final width, but the sharpness of the profile was such that this effect was assumed to be negligible. Also, the irregularities seen in the 2D pinhole image demonstrate a pattern with characteristic size of about 20 pixels; in the calculation of the uncertainty of the 85\% and 15\% brightness levels, the standard deviation of the mean was calculated by dividing the calculated standard deviation at a given radius by the square root of the effective number of uncorrelated pixels at that radius (that is, number of pixels at that radius divided by 20).

\section{Stresses}\label{stresses}

It is informative to examine potential sources of FRD due to the fiber itself. Ideally, a perfect fiber would transmit the exact same signal out as was input, but imperfections in the fiber geometry can induce coupling to higher-order modes and ``spread out'' the input light causing FRD. In this subsection, we investigate three sources of stress on a fiber and their effect on FRD using an effective f/2.8 beam. Effects from bending, twisting, and squeezing are presented and related by calculating stress $\sigma$ on the fiber, with subscripts differentiating stress and FRD. However, only a perfunctory attempt has been made in this work to model FRD as a function of stress, because length over which stress is applied, degree of imperfections in the core/clad interface, fiber cladding Young's modulus, direction of stress applied, and many other factors complicate such an analysis.

\begin{figure}[!htb]
\center{\includegraphics[width=16cm]
{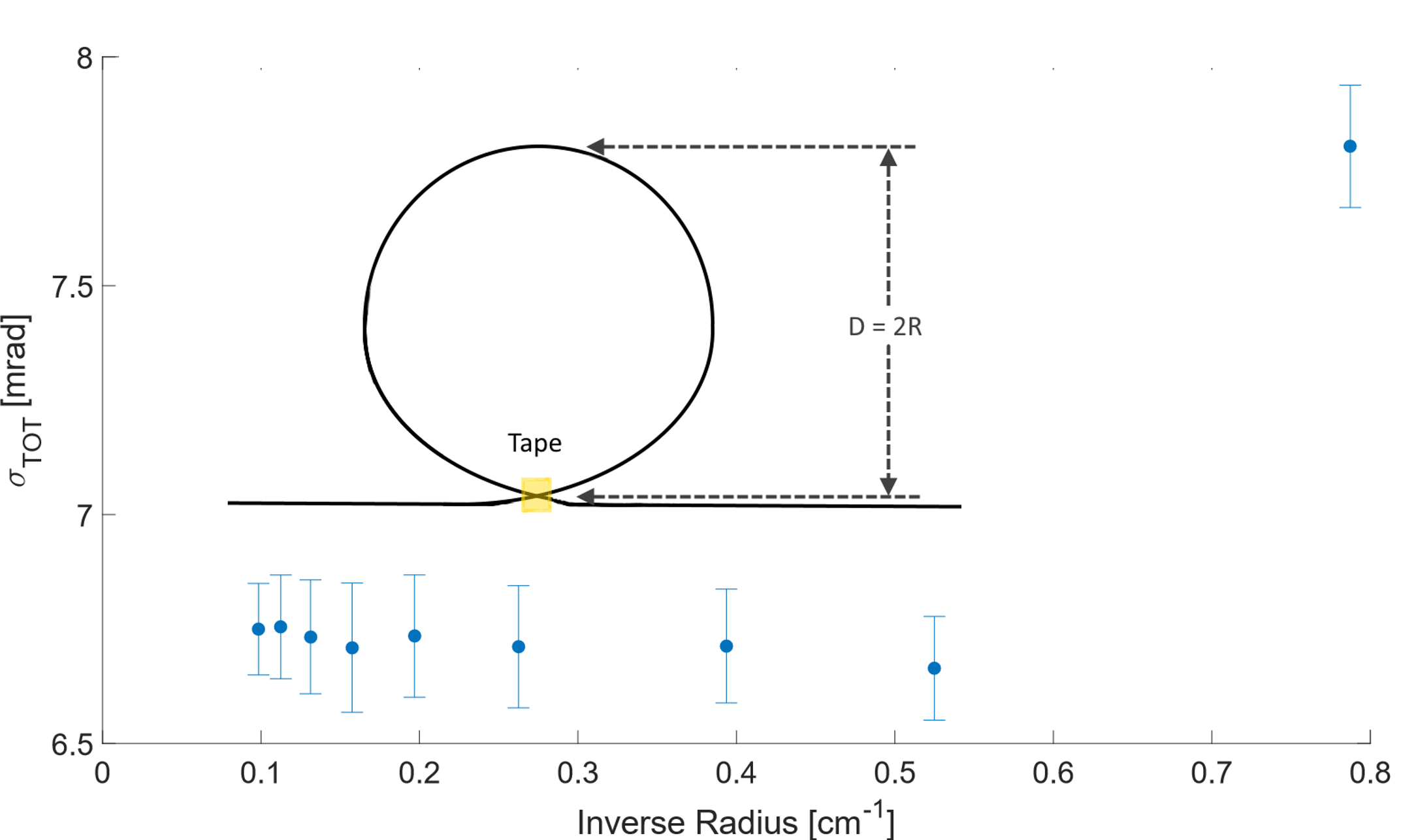}}
\caption{\label{fig:Bending} FRD as a function of bend curvature, with fiber bent as shown in the inset. No observable change in FRD occurred until a bend radius of 1.3 cm.
}
\end{figure}

We first examined stress due to bending. A Polymicro fiber was subject to a loop of a measured, decreasing radius while measuring FRD, with FRD results in Figure~\ref{fig:Bending} and example loop inset. As the literature indicates \cite{Clayton1989}, large-scale bending (with respect to the 127$\mu m$ core of the fiber) does not significantly contribute to the FRD of this system. In particular, no FRD change was found until a bend radius of about $1.25$ cm, where it increased by 1 mrad.

The stress due to a loop of radius $R$ and a fiber with fiberglass core/clad of core radius $r_{core}$ depends linearly on its Young's modulus $E$, as shown in Equation~\eqref{eq:2}. Note that the Young's modulus for fiberglass used is large relative to other materials but standard for optical fibers, while fiber radius (with buffer included) is taken to be small relative to loop radius.

\begin{equation}
\sigma_{bend} \approx Er_{core}\frac{1}{R}\label{eq:2}
\end{equation} \cite{Hefferon87} where $\sigma_{bend}$ corresponds to the induced stress on the fiber. In particular, in this fiber the stress where FRD notably increased would be about 500 MPa as a result.

It is worth considering twist as a potential contribution to FRD in comparison to bending, because twisting is expected to be present in the Cobra fiber positioner system up to a full $360^\circ$ range of motion. It would be informative to determine the stress on the fiber due to twist to predict if there will be an observable effect on FRD. With shear modulus $G$ of clad radius $r_{clad}$, Equation~\eqref{eq:twist} applies for twist of angle $\theta=\pi$

\begin{equation}
\sigma_{twist} = Gr_{clad}\frac{\theta}{L}\label{eq:twist}
\end{equation} \cite{Hefferon87} over length $L$ dictated by the Cobra system, notably similar to the bending stress equation (Equation~\eqref{eq:2}). In this experiment, the fiber is held in place with epoxy at the fiber arm (see Figure~\ref{fig:Cobra}) at the end of the fiber positioner on one side (see Section~\ref{Cobra}), but is not otherwise significantly constrained (allowing $L$ to be on the order of a meter). Thus, the fiber is able to twist over the order of the length of the fiber, and the twist stress is proportionally orders of magnitude lower than that considered in the bend test above. It is expected that twist will not have a significant effect on FRD in this system. In the PFS, the distance between the fiber arm and a segmented tube over which twisting occurs is shorter at about 28 cm, so twist will contribute a larger stress than that from an unconstrained fiber but still significantly smaller than the maximum bending stress tested in this paper.

Beyond bending and twisting that may be induced in fibers during operation of the Subaru telescope, pinching is another possible additional sources of stress. There are strain relief boxes to alleviate this kind of issue, but it is still expected that some amount of applied stress will change during system operation.

To test the effect of pinching on a fiber, a Polymicro fiber was subject to an increasing weight on an aluminum base over a $L=1.5$ cm length of fiber, up to a few kilograms. The weight can be converted to an average pressure after calculating the effect of compression of the fiber itself due to the weight. In particular, the half-width $b$ of the fiber due to the compression is given by

\begin{equation}
b=\sqrt{\frac{4F\left(\frac{1-\nu_1^2}{E_1}+\frac{1-\nu_2^2}{E_2}\right)}{\pi L\left(\frac{1}{R_1}+\frac{1}{R_2}\right)}}\label{eq:3}
\end{equation} \cite{Bamberg06} where $F$ is the force applied to the fiber, $\nu$ is the Poisson's ratio for the two interfacing objects, denoted by the subscripts (let object 1 be the aluminum base and object 2 be the polyimide buffer of the fiber), $E$ is Young's modulus, and $R$ is the radius of each object.

\begin{figure}[!htb]
\center{\includegraphics[width=16cm]
{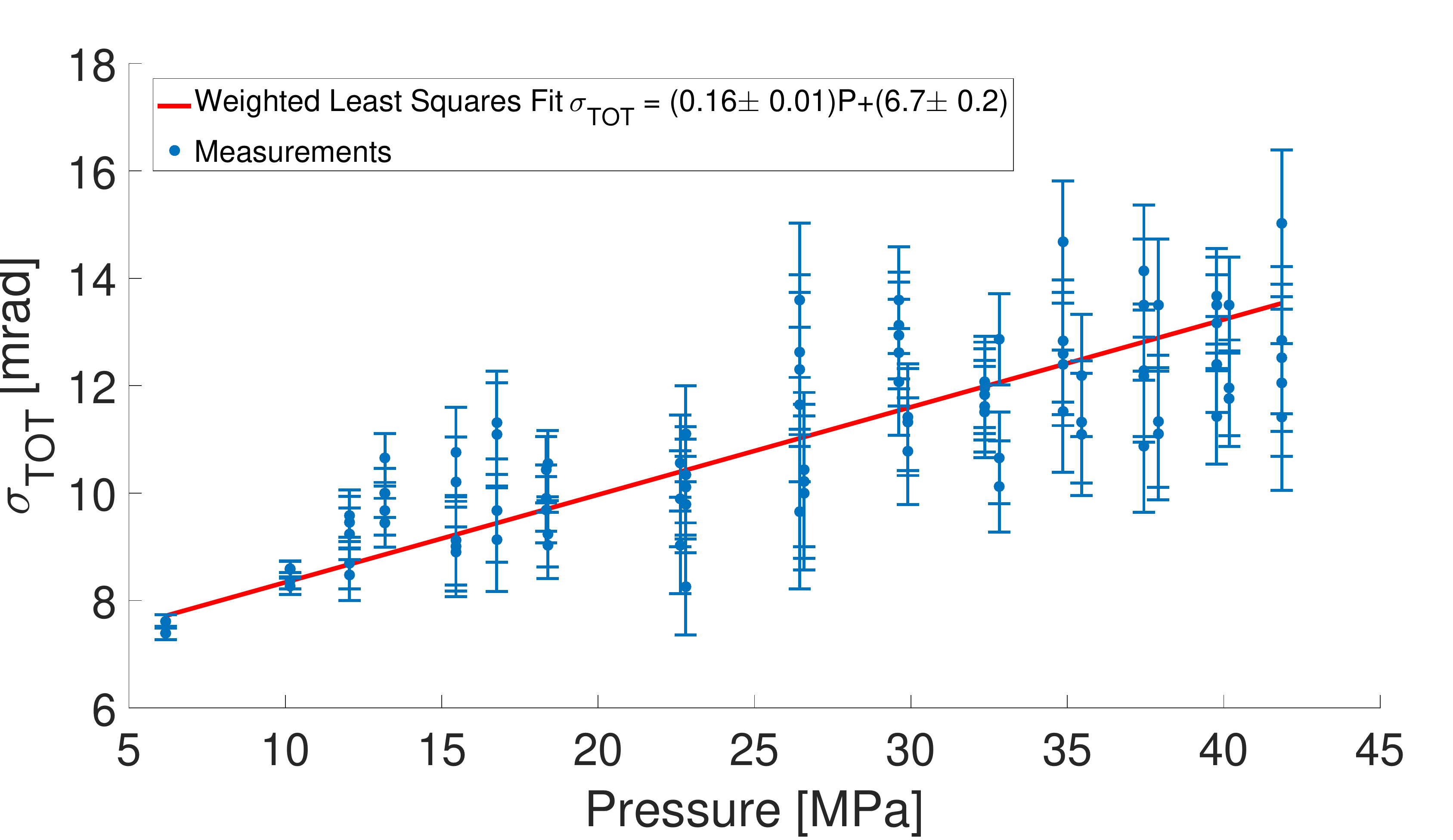}}
\caption{\label{fig:Pressure} Variation of FRD with applied radial pressure to the fiber buffer. FRD appears to grow roughly linearly with applied buffer pressure, although extrapolation to lower pressures may not be accurate. Scatter also increases with applied weight, indicating how small variations in pressure distribution can significantly affect FRD measurements as weight increases. Error bars in this figure indicate the standard deviation of all measurements at a given pressure. 
}
\end{figure}

Average pressure $P$ on the buffer thus is proportional to $\sqrt{F}$. A plot demonstrating the $\sigma_{TOT}$ dependence on $P$ is shown in Figure~\ref{fig:Pressure}. There is an approximately linear relationship between pressure on the buffer and $\sigma_{TOT}$. However, light propagates through the core and depends on the core/clad interface, so the buffer pressure alone may not characterize the physical nature of this FRD increase.

\section{Angular misalignment}\label{AngMis}
Angular misalignment is known to contribute to the power distribution fiber output. Due to this, measurements of FRD often rely on optical systems that are carefully aligned. This holds especially true for the cone test. While ideally in practice every fiber would be coaxial with a uniform beam, in practice, a fiber in a Cobra fiber positioner will move through a range of angular misalignments during operation due to impossibility of perfect angular alignment. Furthermore, due to the design of fibers throughout the focal plane, some fibers necessarily receive a non rotationally symmetric profile, violating the simplification that permits $\sigma_{FRD}$ to characterize the point spread function without additional parameters. Thus, angular misalignments' effects on FRD must be well understood to correct for foreground sky signal throughout telescope operation for any fiber positioning.

\begin{figure}[!htb]
\center{\includegraphics[width=16cm]
{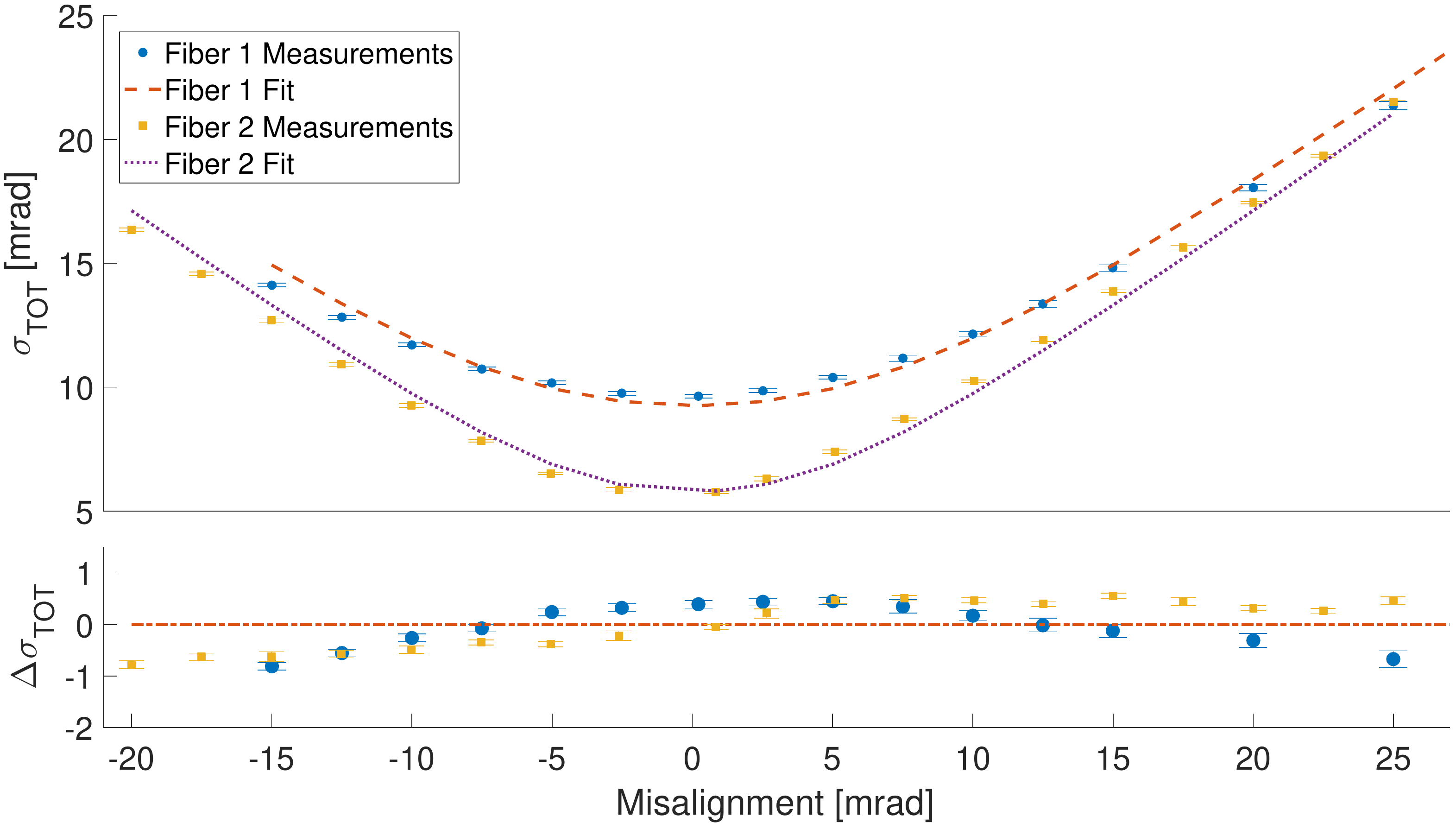}}
\caption{\label{fig:Angular} Variation of FRD with geometric angular misalignment from fitted minimum for two fibers. The effect is approximately hyperbolic, fitted for $\sigma_{FRD}$ (see Section~\ref{AngMis}) and an x-axis offset. The bottom panel shows the residuals, which are subtracted from the fit. The minimum $\sigma_{TOT}$, roughly corresponding to the $\sigma_{FRD}$ of each fiber, varies between the two but is within expected variation.
While the fit is good for just effectively fitting to a minimum, there appears to be a nontrivial residual for the fiber of larger $\sigma_{FRD}$.
}
\end{figure}

A test was performed with a f/2.8 beam using two Polymicro fibers (with ferrules on both ends) to determine the effect of angular misalignment on the measurement of FRD. To investigate the effects of angular misalignment, the FRD from each fiber (taken individually) was minimized with respect to the two angular degrees of freedom in a Newport LPV-1 mount. The minimum was taken to correspond to the least angular misalignment. Then, adjustments to this position were taken in $\approx \frac{1}{8}$ mm steps in the screws holding the mount's angular position (corresponding to about 3 milliradian angular adjustments) with FRD measured at each stop. The angular misalignment was determined from the geometry of the mount relative to its calculated minimum. The results from this test as well as fits to equation~\eqref{eq:JimAngle}'s prediction are shown in Figure~\ref{fig:Angular}. 

\begin{figure}[!htb]
\center{\includegraphics[width=16cm]
{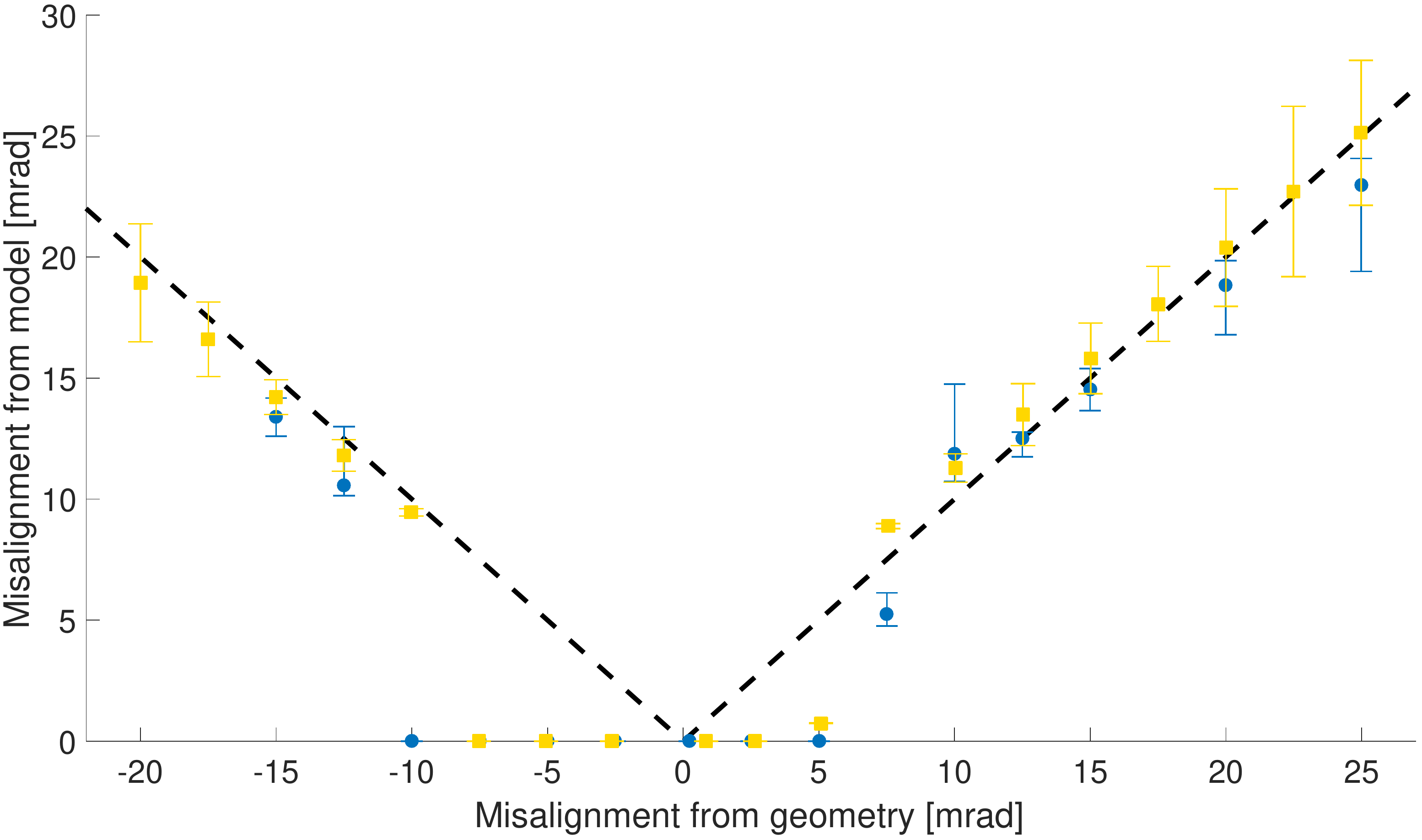}}
\caption{\label{fig:AngularCompare} Angular misalignment between the geometric misalignment and the algorithmically calculated misalignment, with errors calculated from $\sigma_{TOT}$ and the range of angular misalignment with sum of least squares within 5\% of the minimum fit. Agreement between the angles is low at lower angular misalignments, possibly due to nonuniform profile effects that dominate misalignment effects for low misalignment, 
but is good at large angles. This result indicates that the misalignment extraction from the model is reasonable.
}
\end{figure}

However, angular misalignment can also be determined directly from the profile. By taking the width between the 85\% and 15\% brightness positions, a family of 1D profiles with angular misalignment ratios $A$ from 0 to 6 were calculated with 0.1 step increments (See section~\ref{AngMis} for details about $A$) and fitted to the profile, with the final value of $A$ selected from the best fit. Because both $A$ and $\sigma_{TOT}$ are calculated from this model, using equation~\eqref{eq:JimAngle}, $\sigma_{FRD}$ and thus angular misalignment $a$ can also be found. The geometric angle can be compared to the angle predicted from the model to determine the model's efficacy, as shown in Figure~\ref{fig:AngularCompare}

Verification of the angular misalignment effect on FRD with angle is particularly valuable when interpreting the FRD with Cobra fiber position, because an imperfectly aligned Cobra will traverse a range of angular misalignments that can noticeably affect FRD measurements. 

\section{Cobra FRD and Discussion}\label{Cobra}

\begin{figure}[!htb]
\center{\includegraphics[width=16cm]
{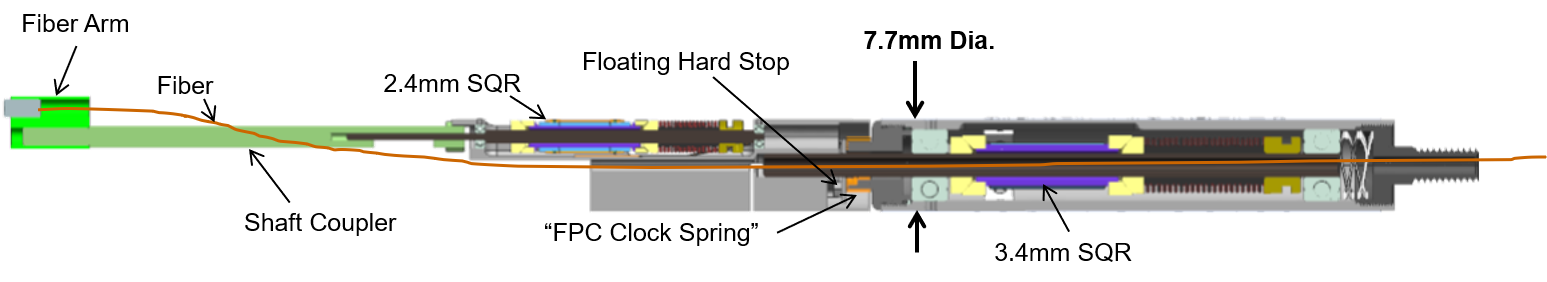}}
\caption{\label{fig:Cobra} Schematic cutaway of a Cobra fiber positioner, with various parts and parameters marked. Figure from the Subaru PFS collaboration's Cobra Fiber Positioner Manufacturing Readiness Review. Also see \citet{Fisher2014} for a Cobra figure and more discussion of the Cobra.
}
\end{figure}

The Cobra Fiber Positioner is a $\theta$-$\phi$ eccentric-axis fiber positioner that permits a fiber to be located anywhere within a 9.5mm patrol region. The Cobra units were constructed by New Scale Technologies, and the positioning of a Cobra was controlled using New Scale Technology's Pathway program with pulses of specified number of steps and length. However, mapping from steps to angular position is not perfectly accurate which does cause an uncertainty in position that is not calibrated for. To minimize angular error, the Cobra fiber positioner was mounted to a machined Thorlabs LCP01 piece to be aligned using the Thorlabs cage system. Figure~\ref{fig:Cobra} depicts a Cobra used in the conducted experiment.

\begin{figure}[!htb]
\center{\includegraphics[width=16cm]
{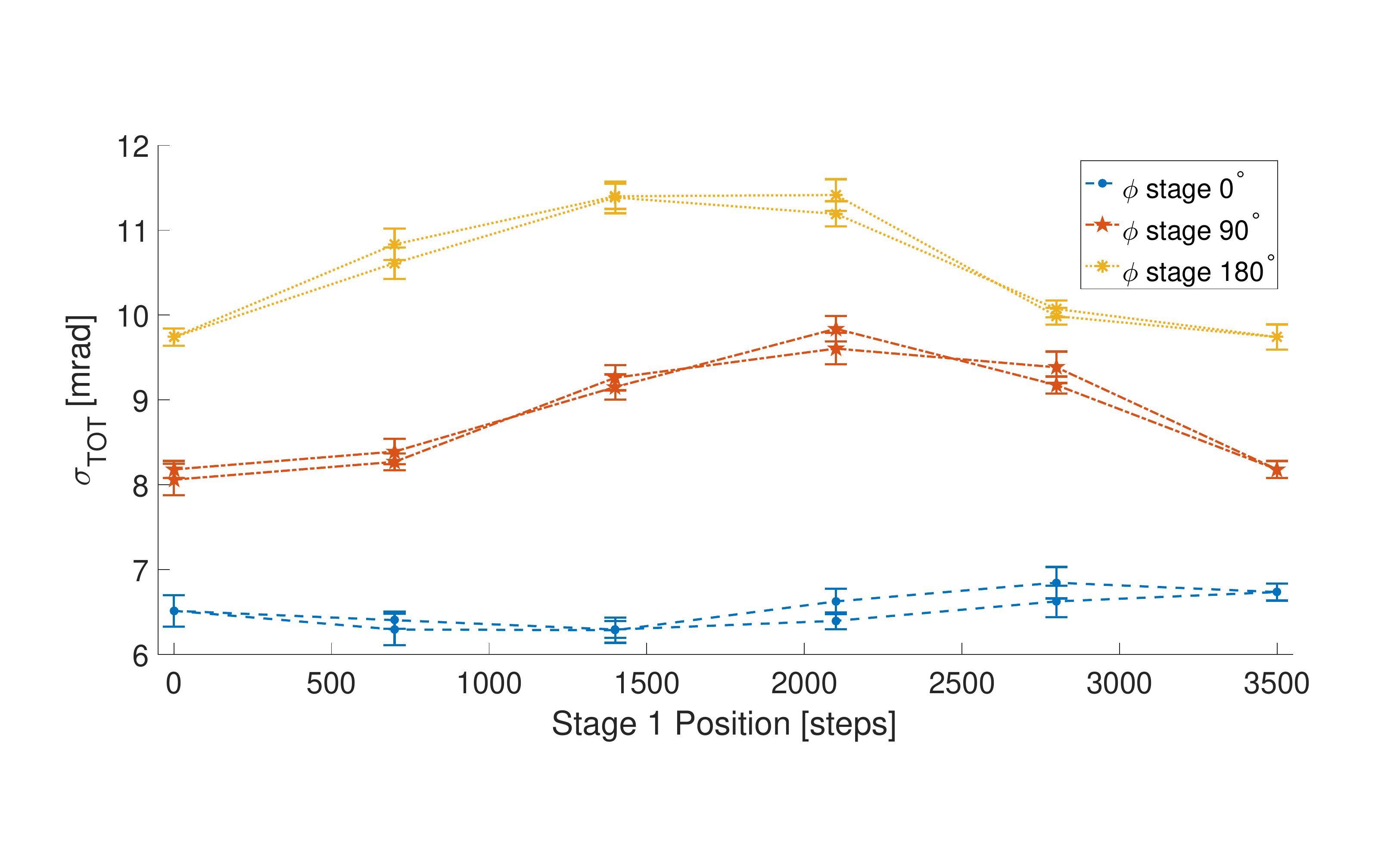}}
\caption{\label{fig:Cobra_SigTOT} Profile $\sigma_{TOT}$ of a fiber with microlens throughout the range of motion in both stages of the Cobra positioner. There is a significant variation of $\sigma_{TOT}$ through stage 1 ($\theta$ stage) and stage 2 ($\phi$ stage) motion. The sinusoidal behaviour across stage 1 indicates an angular misalignment effect. Uncertainties of about $\pm 50$ steps in stage 1, corresponding to $\pm 5^\circ$ angular precision, were also present though not depicted in this figure.
}
\end{figure}

The FRD of a fiber fitted in a Cobra fiber positioner with a microlens was measured throughout the range of the positioner's motion in both stages with a f/2.2 beam, which becomes f/2.8 after the microlens.  Due to the nature of the Cobra mount used, angular alignment could not be adjusted to be minimized. The FRD was measured in steps of 500 steps from 0 to 3500 in stage 1 and back again to determine hysteresis. The measured FRD is shown in Figure~\ref{fig:Cobra_SigTOT}.

The FRD through the positionings in stage 1 ($\theta$ stage) appears to increase and then decrease, completing a cycle in the $360^\circ$ range of stage 1. Furthermore, while FRD varies across stage 1 motion, there is no noticeable systematic increase in FRD after the full $360^\circ$ twist is imposed on the fiber, which would increase twist while minimizing net angular misalignment change. Thus, twist appears to not contribute noticeably to FRD as was expected due to its low imposed stress (see Section~\ref{stresses}). Thus, FRD variations across stage 1 appear to be due to angular misalignment.

Across positions in stage 2, there is a more significant increase in $\sigma_{TOT}$. However, the amplitude of the variations across stage 1 also increase, indicating a larger angular misalignment. This complicates the analysis of bend stress vs. angular misalignment as the primary cause of the FRD increase.

The expected bend stress induced due to the Cobra fiber positioner can be calculated from Equation~\eqref{eq:2}. By substituting in $R = 3.12$ cm, the minimum inverse curvature of the fiber in a Cobra after approximating its path as a helix, the resulting stress is lower than the stress in the bending test where $\sigma_{TOT}$ notably increased.

\begin{figure}[!htb]
\center{\includegraphics[width=16cm]
{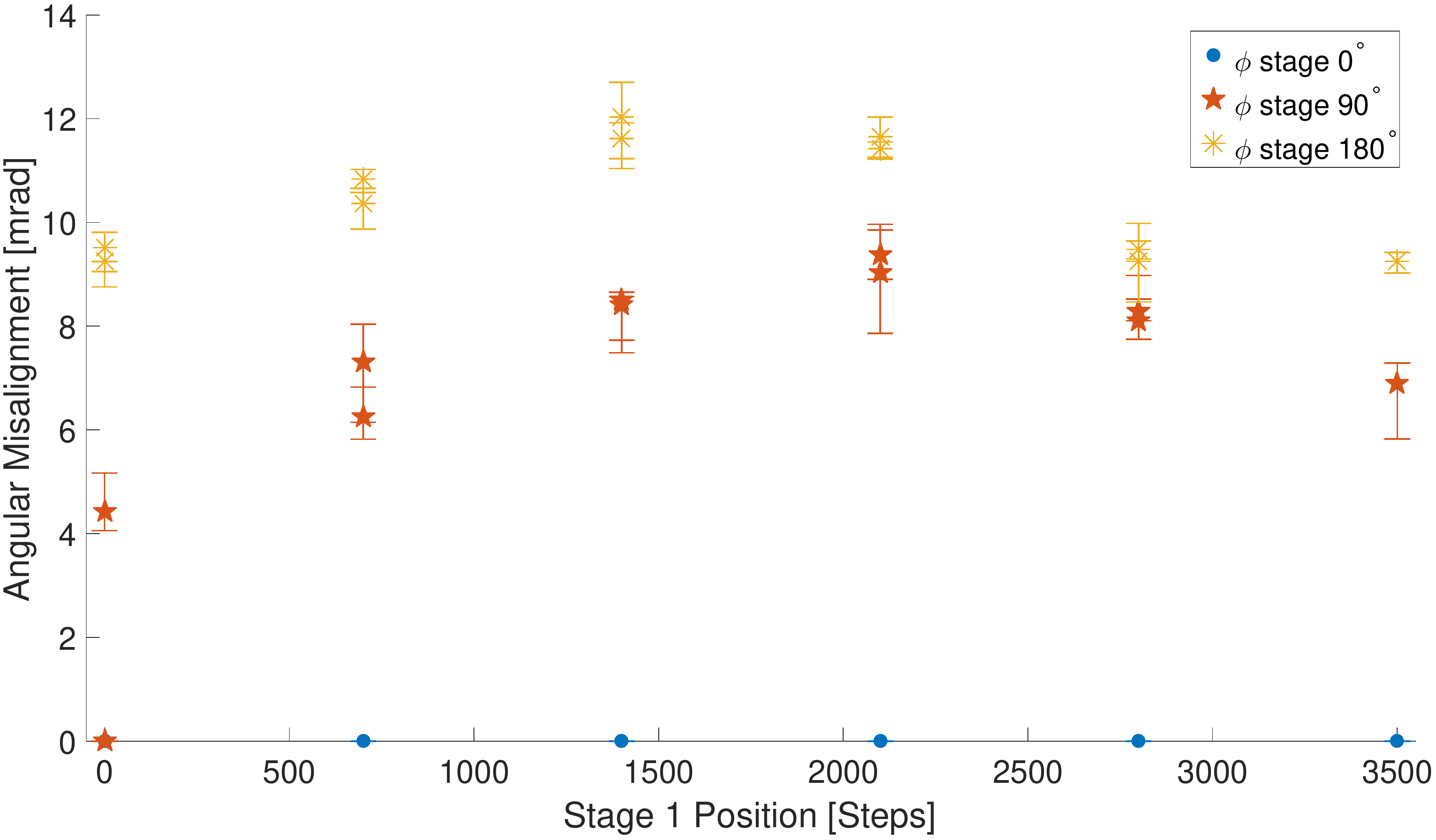}}
\caption{\label{fig:CobraCompare} Angular misalignment as calculated from the power distribution model in this paper (section~\ref{Model}). The sinusoidal behavior across stage 1 suggests that the $\sigma_{TOT}$ variation in stage 1 is indeed due to angular misalignment rather than stress, and angular misalignment due to the $\phi$ stage is also apparent.
}
\end{figure}

The angular misalignment can be measured by applying Equation~\eqref{eq:JimAngle} to the range of FRDs through stage 1 or stage 2 motion and calculating the variation of angular misalignments required to generate the resulting plot. This angular misalignment measurement can help give insight about the possible effect on FRD during Cobra operation. The extracted angular data agrees with the above predictions, as seen in Figure~\ref{fig:CobraCompare}.

\begin{figure}[!htb]
\center{\includegraphics[width=16cm]
{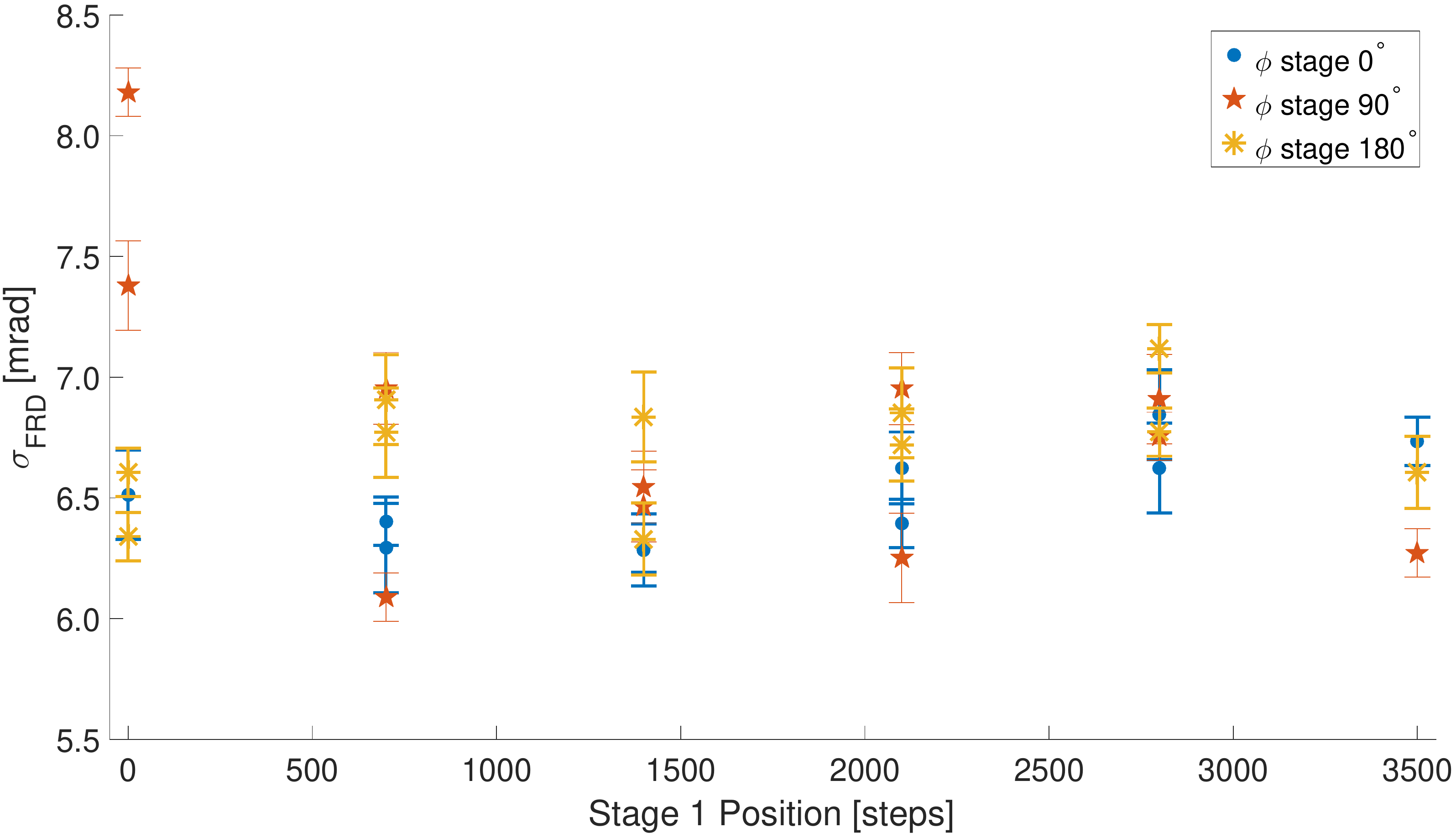}}
\caption{\label{fig:CobraFRD} FRD extracted from the fiber profiles after considering angular misalignment. The FRD can be well-characterized in a single experimental setup across varying configurations due to this angular misalignment extraction, and is consistent with minimal stress on the fiber during Cobra operation.
}
\end{figure}

The angular misalignment extracted from the profiles thus allows for $\sigma_{FRD}$ to be extracted directly from the profiles in this system, as shown in Figure~\ref{fig:CobraFRD}. These values of $\sigma_{FRD}$ are broadly consistent with a constant $\sigma_{FRD}$ of about 6.5 mrad throughout all fiber positions, indicating that the stresses during operation did not significantly increase FRD. From the values of $\sigma_{TOT}$ alone, it would be impossible to determine the source of profile broadening.

In conclusion, a model of fiber profile due to FRD and angular misalignment was presented and used to analyze cone profiles of a fiber under stress or angular misalignment. While high stresses were found to increase FRD, FRD did not significantly increase for low stress. Angular misalignment was found to have a significant effect on the profile width, which was fitted by the model considered. The FRD component of the model is represented as a Gaussian as seen in previous literature results (e.g. \citet{Haynes2011}), whereas the angular misalignment component of the profile is simplified into one parameter. Some stresses that would be present in the Subaru PFS's Cobra fiber positioner system are considered and expected to be low during operation. The angular misalignment from the presented model is compared to a test with angular misalignment calculated geometrically and found to be consistent for angular misalignments larger than about 5 milliradians. Thus, the model could then be applied to the Cobra fiber positioner throughout its range of motion, indicating minimal FRD increase, if any, that would have otherwise been obscured by the angular misalignment present in the system. An experimental setup such as used in this paper combined with the model used should simplify analysis of FRD in dynamic systems without the need to realign for each configuration to be tested.

\section*{Acknowledgments}

We are grateful to the Kavli Institute for the Physics and Mathematics of the Universe, the National Astronomical Observatory of Japan, and the Subaru PFS collaboration for their feedback and recommendations throughout this project. We would especially like to thank Naoyuki Tamura and Yuki Moritani for discussions and direction for this project, as well as Leandro dos Santos for fiber preparation.  This material is based upon work supported by the National Science Foundation under MSIP grant no. 1636426. 

This research has made use of NASA's Astrophysics Data System.
 
\bibliographystyle{ws-jai}
\bibliography{references}

\end{document}